\documentclass[preprint, 11pt]{elsarticle}

\usepackage{setspace}

\usepackage[export]{adjustbox}
\usepackage{amssymb}
\usepackage{amsthm}
\usepackage{amsmath}
\usepackage{booktabs}
\usepackage{fontawesome5}
\usepackage[bookmarks=false]{hyperref}
\hypersetup{colorlinks, linkcolor=blue, citecolor=blue, urlcolor=blue}
\usepackage{lineno}
\usepackage[figuresright]{rotating}
\usepackage{soul}
\usepackage{stfloats}
\usepackage{tikz}
\usepackage[colorinlistoftodos,prependcaption,textsize=tiny]{todonotes}
\usetikzlibrary{shapes.geometric, arrows, positioning, backgrounds}
\usepackage{xargs}
\usepackage[pdftex,dvipsnames]{xcolor}
\usepackage[margin=1in]{geometry}

\biboptions{sort&compress}

\renewcommand{\hl}[1]{#1}

\newcommandx{\unsure}[2][1=]{\todo[linecolor=red,backgroundcolor=red!25,bordercolor=red,#1]{#2}}
\newcommandx{\change}[2][1=]{\todo[linecolor=blue,backgroundcolor=blue!25,bordercolor=blue,#1]{#2}}
\newcommandx{\info}[2][1=]{\todo[linecolor=OliveGreen,backgroundcolor=OliveGreen!25,bordercolor=OliveGreen,#1]{#2}}
\newcommandx{\improvement}[2][1=]{\todo[linecolor=Plum,backgroundcolor=Plum!25,bordercolor=Plum,#1]{#2}}

\begin{document}

\doublespacing

\begin{frontmatter}

\title{A Multimodal Manufacturing Safety Chatbot: Knowledge Base Design, Benchmark Development, and Evaluation of Multiple RAG Approaches}
% Other titles to consider:
%% (a) Multi‑Method RAG for Manufacturing Safety: System Design and Comparative Evaluation
%% (b) Evaluating Retrieval‑Augmented Generation Pipelines for Manufacturing Safety Guidance

\author[a]{Ryan Singh} 
\author[b]{Austin Hamilton}
\author[a]{Amanda White}
\author[c,d]{Michael Wise}
\author[c]{Ibrahim Yousif}
\author[a]{Arthur Carvalho}
\author[a]{Zhe Shan}
\author[c]{Reza Abrisham Baf}
\author[c]{Mohammad Mayyas}
\author[e]{Lora A. Cavuoto}
\author[a]{Fadel M. Megahed\corref{cor1}}
\ead{fmegahed@miamioh.edu}

\address[a]{Farmer School of Business, Miami University, 800 E. High Street, Oxford, OH 45056, USA}
\address[b]{Department of Computer Science and Software Engineering, Miami University, 105 Tallawanda Road, Oxford, OH 45056, USA}
\address[c]{Department of Engineering Technology, Miami University, 1601 University Blvd., Hamilton, OH 45011, USA}
\address[d]{Department of Mechanical and Manufacturing Engineering, Miami University, 650 E. High Street, Oxford, OH 45056, USA}
\address[e]{Department of Industrial and Systems Engineering, University at Buffalo, 407 Bell Hall, Buffalo, NY, 14260}

\begin{abstract}
Ensuring worker safety remains a critical challenge in modern manufacturing environments. Industry 5.0 reorients the prevailing manufacturing paradigm toward more human-centric operations. Using a design science research methodology, we identify three essential requirements for next-generation safety training systems: high accuracy, low latency, and low cost. 
We introduce a multimodal chatbot powered by large language models that meets these design requirements. The chatbot uses retrieval-augmented generation to ground its responses in curated regulatory and technical documentation. To evaluate our solution, we developed a domain-specific benchmark of expert-validated question and answer pairs for three representative machines: a Bridgeport manual mill, a Haas TL-1 CNC lathe, and a Universal Robots UR5e collaborative robot. We tested 24 RAG configurations using a full-factorial design and assessed them with automated evaluations of correctness, latency, and cost. Our top 2 configurations were then evaluated by ten industry experts and academic researchers. Our results show that retrieval strategy and model configuration have a significant impact on performance. The top configuration, selected for chatbot deployment, achieved an accuracy of 86.66\%, an average cost of \$0.005 per query, and \hl{an average end-to-end latency of 10.04 seconds. This latency is practical for delivering a complete safety instruction and is measured from query submission to full instruction delivery rather than generation onset.} Overall, our work provides three contributions: an open-source, domain-grounded safety training chatbot; a validated benchmark for evaluating AI-assisted safety instruction; and a systematic methodology for designing and assessing AI-enabled instructional and immersive safety training systems for Industry 5.0 environments.
\end{abstract}

\cortext[cor1]{Corresponding author. Tel.: +1-513-529-4185.}
\end{frontmatter}

% \linenumbers
\section{Introduction}
\label{sec:intro}
Manufacturing continues to play a central role in the U.S. economy. In 2024, it generated more than \$2.9 trillion in annual GDP, or roughly 10\% of total U.S. economic output, according to the Bureau of Economic Analysis \citep{bea2024manufacturing}. The sector employs over 12.7 million workers \citep{bea2025employment} and drives around 54\% of the research and development conducted by American businesses with 10 or more domestic employees in 2021 \citep{nsf2024rd}. These figures highlight both the economic importance of the sector and the need to support the safety and productivity of its workforce.

During the past 15 years, manufacturing has experienced a significant shift with the rise of smart manufacturing technologies, often referred to as Industry 4.0. These advances include the Industrial Internet of Things (IIoT), cyber-physical systems (CPS), collaborative robots (cobots), machine learning (ML), and other digital tools that provide real-time data for process monitoring and system adaptation \citep{xu2021industry}. Even with these improvements, Industry 4.0 still faces limitations. Many systems operate close to their technological ceiling, where adding more automation produces only small gains. Legacy equipment also makes it difficult to achieve interoperability and reconfigure systems quickly. As production demands change and human robot interaction becomes more common, issues such as ergonomic strain, operator fatigue, and safety concerns continue to appear. These ongoing challenges show that productivity gains alone are not enough to ensure resilience. More recently, Industry 5.0 has emerged to address these concerns by placing human centricity, resilience, and sustainability on equal footing with efficiency \citep{xu2021industry,lu2022outlook}. By encouraging closer collaboration between humans and intelligent machines, Industry 5.0 points toward manufacturing systems that are more adaptable, safer, and better able to recover from disruptions while continuing to improve through digital intelligence.

Despite its economic importance, manufacturing continues to face persistent safety challenges. According to the U.S. Bureau of Labor Statistics \citep{bls2024injuries}, the sector reported over 320,000 workplace injuries, making it one of the most hazardous occupational domains.
\hl{Common incidents include being struck by objects or equipment (3$^{\text{rd}}$ most common workplace injury) and being caught in or compressed by equipment/ objects (4$^{\text{th}}$ most common)}\cite{LibertyMutual2025WSI}.
These events endanger workers and impose substantial economic costs on employers and the broader economy. The manufacturing sector spends $\sim$\$7.5 billion annually on direct medical payments and lost wages \citep{LibertyMutual2025WSI}. Non-fatal workplace injuries result in an average of 11 lost workdays per employee, equivalent to $\sim$\$1,590 in lost productivity per case \citep{peterson2021average}.

Contributing factors to workplace safety incidents include systems and worker-level factors. Along with the traditional safety concerns, the evolving manufacturing environment has created new challenges for manufacturing operators, as they need to know about and interact with new systems. For example, variations in equipment make it difficult for operators to remember operating procedures across a range of systems. When this is accompanied by the time pressure imposed by high production demands, safety incidents may be more likely to occur. 

This is where the application of AI can have a positive influence. 
For example, prior studies have identify applications of AI in manufacturing to support operators in performing their tasks, including the presentation of maintenance guidelines, process monitoring, predicting machine downtime, and worker training \citep{kiangala2024experimental,mayyas2020fenceless}. Modernizing worker training is a critical area for advancing worker performance and mitigating safety risk. Traditional training has often involved classroom or online lecture-based content presented upon job entry and/or at fixed intervals (e.g., annually). This approach is static and fails to reflect the complexity of modern manufacturing environments or worker experiences. In this static setup, when an operator needs information about a piece of equipment, they may not be able to easily access information in a user manual or online. There is thus a pressing need for adaptive, context-aware training systems capable of providing personalized, accurate, and readily accessible safety information in real time.

Following a design science research (DSR) methodology \citep{peffers2007design}, our work identifies three essential requirements that a next-generation manufacturing safety training solution must satisfy: high accuracy, low cost, and low latency. To address these requirements, we propose an open-source, Retrieval Augmented Generation \hl{(RAG)} based chatbot grounded in regulatory documents such as OSHA standards and OEM manuals. Prior work in construction safety shows that chatbot-based training can improve hazard awareness, especially for participants with limited on-site experience  \citep{zhu2022can}. One advantage identified in that work is the ability to tailor training materials to specific worker needs. By combining large language models (LLMs) with curated, domain-specific sources, our RAG system provides precise and contextually relevant safety guidance while maintaining affordability and responsiveness.

A second contribution of our work is the development of a domain-specific benchmark that evaluates the chatbot’s safety reasoning capabilities. Our benchmark spans three generations of manufacturing technology. It includes a legacy manually operated system represented by the \textit{Bridgeport Manual Mill}, an automated Industry 3.0 system represented by the \textit{TL-1 CNC} machine, and an Industry 5.0 collaborative system represented by the \textit{Universal Collaborative Robot} (Cobot). We constructed datasets of question and answer pairs related to each machine. All datasets were vetted by industry practitioners and researchers with expertise in smart manufacturing, industrial engineering, occupational safety, and human factors. These benchmark datasets support reproducible evaluation of accuracy and latency and form a foundation for future research in safety-oriented conversational AI systems.

In summary, our work contributes to the growing literature on AI for occupational safety and training by showing how RAG systems can transform static safety instruction into a dynamic/interactive experience. By integrating regulatory grounding, multimodal reasoning, and open benchmarking, our research offers a pathway for using AI to enhance worker safety, compliance, and preparedness in \textit{Industry 5.0} environments.

\section{Literature Review}
\label{sec:litreview}
Existing research in manufacturing has explored a range of AI-driven approaches to enhance worker safety, hazard awareness, and on-site decision support. Three interrelated themes emerge from this work. The first is the use of AI, including LLMs and RAG, for training and decision-making. The second is the development of multimodal interfaces, such as vision and language systems, to support richer human-machine interaction. The third is the use of digital twin environments for simulation, monitoring, and operator support. This section reviews representative work in each of these areas.

Conversational AI frameworks have begun to help operators interpret real-time industrial data. For example, Jeon et al. \cite{jeon2025chatcnc} observe that limited digital literacy among shop-floor workers often makes conventional monitoring interfaces difficult to use. Their system, \textit{ChatCNC}, addresses this issue by allowing operators to ask plain-language questions about machine conditions, such as whether a machine is overheating, instead of navigating complex software menus and interfaces.

Research in this space reflects a broader shift toward AI-mediated instruction and monitoring that aims to reduce human error and improve safety outcomes. Natural language techniques such as LLMs are particularly attractive because they blend general reasoning capability with a conversational interface that aligns well with Industry 5.0’s emphasis on human-centric smart manufacturing. For example, Wang et al. \cite{wang2024llm} developed a collaborative robot with vision–language abilities that enable it to visually map its surroundings and follow spoken or written instructions through an LLM-based parser. Similarly, Lou et al. \cite{lou2025large} built an LLM-powered cognitive agent that plans and makes decisions for assembly and disassembly tasks. Their agent demonstrates context awareness and outperforms traditional automation by leveraging the LLM’s extensive knowledge and reasoning abilities. These examples show how LLMs can function as conversational intermediaries that translate operator intentions into complex technical actions.

A major challenge for LLMs in manufacturing is the gap between broad model training and the specific, real-time knowledge needed on the factory floor. RAG has emerged as a practical solution to this issue by pairing an LLM with a retrieval component that supplies relevant domain information such as documents, sensor readings, or knowledge base entries. Fan et al. \cite{fan2025MaViLa} illustrate this approach with \texttt{MaViLa}, a vision–language model capable of interpreting live visuals and grounding its guidance with retrieved manufacturing knowledge. While \texttt{MaViLa} focuses on broad manufacturing tasks such as process understanding and skill acquisition, our work targets manufacturing safety and uses RAG to ground responses in a curated safety knowledge base.

Another important development in manufacturing involves the growing adoption of multimodal interfaces that integrate vision, language, and other data modalities. Many shop-floor scenarios are inherently visual, such as identifying hazards or verifying proper use of protective equipment, which means an AI assistant that can both see and converse offers clear advantages over text-only systems. The industrial metaverse described by Li et al. \cite{li2024industrial} envisions manufacturing environments where workers and AI interact through virtual and augmented reality tools that combine digital avatars, spatial audio, mixed-reality visuals, and text. These trends demonstrate that an effective manufacturing chatbot must be multimodal so that it can process images or sensor data and generate language that supports safety monitoring and operator assistance. Multimodality therefore serves as a core design principle in our work.

Digital twins and virtual environments have also emerged as key platforms for integrating LLMs, RAG, and multimodal interfaces in realistic, safety-oriented settings. A digital twin is a dynamic virtual representation of a physical system that maintains continuous two-way communication with its real-world counterpart, enabling real-time monitoring, predictive simulation, control, and what-if analysis. In manufacturing, these technologies support proactive safety management and training without interrupting production. Gautam et al. \cite{gautam2025iiot} demonstrate this potential by linking an LLM-based agent system with a factory digital twin. Their system uses specialized LLM agents for machine expertise, data visualization, and fault diagnosis to process streaming IIoT data and reflect the factory state in a virtual model. The digital twin then offers an interactive 3D interface where operators can query an avatar about system status or troubleshooting steps, allowing them to gain insights without approaching hazardous equipment.

Taken together, these developments show the main components of a modern manufacturing safety assistant. They demonstrate that LLMs support natural communication, that RAG provides timely and relevant domain knowledge, and that multimodal understanding is essential for interpreting factory conditions. 
\hl{Despite this progress, the literature still lacks (1) domain-grounded, safety-focused benchmarks with expert-validated gold answers, (2) systematic, controlled comparisons of multiple retrieval strategies and model configurations using accuracy--latency--cost criteria, and (3) a reproducible methodology linking knowledge-base design choices to end-to-end RAG performance for manufacturing safety guidance.
Our work addresses these gaps by proposing a RAG-based chatbot for manufacturing safety training and hazard prevention. Our system uses an extensive safety knowledge base, perceives the manufacturing environment, and engages in dialog to deliver practical safety guidance and real-time risk analysis. Moreover, we contribute an expert-validated benchmark spanning representative manufacturing equipment and describe a systematic methodology for selecting RAG configurations.}

\section{Research Methodology}
\label{sec:methods}
We address the challenge of modernizing hazard-recognition training in manufacturing by adopting a design science research methodology (DSR) \citep{peffers2007design}. This methodological approach is well-suited for the development and evaluation of information technology solutions (artifacts) that aim to solve complex, real-world problems. Following DSR ensures that our work systematically integrates scientific rigor with practical relevance, producing outcomes that are both theoretically grounded and operationally viable within industrial settings.

In particular, our research process follows the six-stage model outlined by Peffers et al. \cite{peffers2007design}, which we elaborate on throughout the paper: (1) problem identification and motivation, (2) definition of solution objectives, (3) artifact design and development, (4) demonstration, (5) evaluation, and (6) communication. We have already identified and motivated the problem in the preceding section, and the last stage of the DSR methodology, i.e., communication, is represented by this paper. That said, we now turn to the second stage, namely defining the solution objectives, which are derived ``\textit{from the problem definition and knowledge of what is possible and feasible}" \citep{peffers2007design}. To operationalize this stage, we established a set of \textit{design requirements} that any proposed artifact must satisfy to address the underlying hazard-recognition and safety-training challenges.

Our first design requirement concerns system \textit{accuracy}, which represents the non-negotiable foundation of any effective safety-training system. In industrial environments, inaccurate or misleading instructions can lead to improper procedures, equipment misuse, and, in the worst cases, fatal accidents. A training artifact that compromises factual precision risks undermining the very purpose of hazard education. Therefore, every element of the system, whether textual guidance, visual cues, or procedural demonstrations, must convey information that is technically correct, operationally safe, and fully aligned with regulatory and manufacturer standards. The following requirement definition captures the above discussion.
%Arthur's note for the future: Design principle can be related to grounding, such as "artifacts must ground their answers on official standards and manufacturer guidelines"
\begin{quote}
    \textbf{Design Requirement \#1 (Accuracy)}: Training artifacts must deliver information that is technically correct and operationally safe.
\end{quote}

Our second design requirement concerns system \textit{latency}, reflecting the need for timely and responsive feedback within dynamic manufacturing environments. In industrial contexts, hazards can emerge and evolve in seconds, and delayed or sluggish responses can render safety guidance ineffective or even dangerous. A training system that fails to deliver instructions, alerts, or assessments in real time risks losing its relevance to the unfolding situation on the shop floor. Therefore, any proposed solution must minimize \hl{end-to-end} latency in both data processing and user feedback, ensuring that workers receive immediate, context-appropriate information as they interact with machinery or simulated environments. \hl{Here, latency is defined as the total time from user query submission to \textit{complete} instruction delivery (i.e., full response rendered), not time-to-first-token or generation onset.} Low \hl{end-to-end} latency is essential not only for maintaining user engagement and situational awareness but also for reinforcing procedural learning under realistic temporal conditions. The following definition summarizes our second design requirement.
\begin{quote}
    \textbf{Design Requirement \#2 (Latency)}:  Artifacts must provide timely and responsive feedback, ensuring that safety instructions are rapidly provided under changing conditions in manufacturing environments.
\end{quote}

%Arthur's note for the future: design principle can be phrased like Cost-efficiency must be embedded in the artifact’s design philosophy, favoring lightweight deployment, open-access frameworks, and the use of existing devices or browser-based systems whenever feasible. 
Our third design requirement concerns the \textit{cost} of deployment and operation, emphasizing the importance of economic accessibility for broad adoption across the manufacturing sector. Safety innovation cannot achieve its intended societal impact if the associated technologies are prohibitively expensive to implement, maintain, or scale. Many small and medium-sized manufacturers, which account for 98 percent of U.S. manufacturing firms \citep{sba2025facts}, face constrained capital and ``do not have the resources'' to invest in advanced technologies and associated training \citep{armstrong2021advanced}. Therefore, training solutions must minimize recurring expenses such as software licensing, hardware requirements, and data-processing costs, without compromising instructional quality or safety integrity. 
\begin{quote}
\textbf{Design Requirement \#3 (Cost)}: Artifacts must minimize deployment and operational costs to ensure affordability and, consequently, adoption across manufacturers of varying sizes.
\end{quote}

\begin{figure*}[htb!]
\centering

\begingroup

\definecolor{phaseA}{HTML}{E8F1FF}
\definecolor{phaseTitle}{HTML}{333333}
\definecolor{stepborder}{HTML}{888888}

\tikzset{
  phase/.style={
  rectangle, draw, ultra thick, rounded corners,
  minimum height=1.5cm, text centered, fill=white,
  text width=5cm, font=\small
  },
  subtask/.style={
  rectangle, draw, very thick, minimum height=1.2cm,
  fill=white, text width=3.25cm, font=\footnotesize
  },
  phaseBox/.style={
    draw=black!15,
    rounded corners=3pt,
    fill=#1,
    inner sep=3pt
  },
  stepBox/.style={
    draw=stepborder,
    fill=white,
    rounded corners=2pt,
    inner sep=3pt,
    align=left
  },
  flowArrow/.style={-latex, very thick, draw=black!35},
  arrow0/.style={ultra thick,->,>=stealth},
  arrow/.style ={very thick,->,>=stealth},
  color1/.style={draw={rgb,255:red,254;green,240;blue,217}},
  color2/.style={draw={rgb,255:red,253;green,204;blue,138}},
  color3/.style={draw={rgb,255:red,252;green,141;blue,89}},
  color4/.style={draw={rgb,255:red,239;green,101;blue,72}},
  color5/.style={draw={rgb,255:red,215;green,48;blue,31}},
  color6/.style={draw={rgb,255:red,179;green,0;blue,0}}
}

\begin{tikzpicture}[framed]

  % Phases
  \node (phase1) [phase, color1] {\faDatabase\, \textbf{Phase 1: Source Curation and Corpus Development}};
  \node (phase2) [phase, color2, right=2.5cm of phase1] {\faProjectDiagram\, \textbf{Phase 2: Knowledge Base and Indexing}};
  \node (phase3) [phase, color3, below=3.5cm of phase2] {\faRobot\, \textbf{Phase 3: Multimodal Chatbot}};
  \node (phase4) [phase, color4, below=3.5cm of phase1] {\faClipboardList\, \textbf{Phase 4: Benchmark Q\&A Dataset Development}};
  \node (phase5) [phase, color5, below=3.5cm of phase4] {\faCogs\, \textbf{Phase 5: Automated Evaluation and Experimental Design}};
  \node (phase6) [phase, color6, below=3.5cm of phase3] {\faTrophy\, \textbf{Phase 6: Human Evaluation}};

  % Phase 1 subtasks
  \node (p1t1) [subtask, below left=0.15cm and -2cm of phase1, color1] {\faBook\, Collect safety documents};
  \node (p1t2) [subtask, below right=0.15cm and -3cm of phase1, color1] {\faFile\, Extract and clean text};
  \node (p1t3) [subtask, below=0.25cm of p1t2, color1] {\faCrop\, Content segmentation};
  \node (p1t4) [subtask, below=0.25cm of p1t1, color1] {\faTags\, Filename standardization};

  % Phase 2 subtasks
  \node (p2t1) [subtask, below left=0.15cm and -3cm of phase2, color2] {\faDatabase\, Cloud-Based keyword and semantic search indices};
  \node (p2t2) [subtask, below right=0.15cm and -2cm of phase2, color2] {\faSortAlphaDown\, Local index for keyword search};
  \node (p2t3) [subtask, below=0.25cm of p2t1, color2] {\faProjectDiagram\, Graph-based indices};
  \node (p2t4) [subtask, below=0.25cm of p2t2, color2] {\faLayerGroup\, Retrieval routing across heterogeneous indices};

% Phase 3 subtasks
\node (p3t1) [subtask, below left=0.15cm and -3cm of phase3, color3]
{\faComments\, Chat UI};
\node (p3t2) [subtask, below right=0.15cm and -2cm of phase3, color3]
{\faImage\, Multimodal LLM for text and vision capabilities};
 \node (p3t3) [subtask, below=0.25cm of p3t2, color3]
{\faMicrophone\, Speech capabilities};
\node (p3t4) [subtask, below=0.25cm of p3t1, color3]
  {\faLaptopCode\, Unified multimodal user interface};

  % Phase 4 subtasks
  \node (p4t1) [subtask, below right=0.15cm and -3cm of phase4, color4]  {\faPen\ Draft safety questions from collected sources};
  \node (p4t2) [subtask, below left=0.15cm and -2cm of phase4, color4] {\faReply\ Write gold answers};
  \node (p4t3) [subtask, below=0.25cm of p4t2, color4] {\faUserCheck\, Validate with industry and safety experts};
  \node (p4t4) [subtask, below=0.25cm of p4t1, color4] {\faBalanceScale\, Finalize reference answers};

  % Phase 4 subtasks
  %\node (p4t1) [subtask, below right=0.15cm and -3cm of phase4, color4] {\faSlidersH\, Define experimental factors and their levels};
  %\node (p4t2) [subtask, below left=0.15cm and -2cm of phase4, color4] {\faRobot\, Generate answers for each configuration};
  %\node (p4t3) [subtask, below=0.25cm of p4t2, color4] {\faClipboardCheck\, Compute automated metrics (7 evaluators)};
  %\node (p4t4) [subtask, below=0.25cm of p4t1, color4] {\faFilter\, Rank and select top $n$ configurations};

  % Phase 4 subtasks
\node (p5t1) [subtask, below right=0.15cm and -3cm of phase5, color5] {\faSlidersH\, Define experimental factors and their levels};
\node (p5t2) [subtask, below left=0.15cm and -2cm of phase5, color5] {\faRobot\, Generate answers for each configuration};
\node (p5t3) [subtask, below=0.25cm of p5t2, color5] {\faClipboardCheck\, Compute automated metrics};
\node (p5t4) [subtask, below=0.25cm of p5t1, color5] {\faFilter\, Rank and select top $n$ configurations};

% Phase 6 subtasks
\node (p6t1) [subtask, below left=0.15cm and -3cm of phase6, color6] {\faTable\, Curate human evaluation Q\&A benchmark set};
\node (p6t2) [subtask, below right=0.15cm and -2cm of phase6, color6] {\faEyeSlash\, Conduct blind human grading of curated set};
\node (p6t3) [subtask, below=0.25cm of p6t2, color6] {\faCalculator\, Aggregate human scores per configuration};
\node (p6t4) [subtask, below=0.25cm of p6t1, color6] {\faTrophy\, Select the final and winning configuration};

  % Phase arrows
  \draw [arrow0] (phase1.east) -- (phase2.west);
  \draw [arrow0] (phase2.east) --  ++(1.75,0) |- (phase3.east);
  \draw [arrow0] (phase3) -- (phase4);
  \draw [arrow0] (phase4.west) -- ++(-1.75,0) |- (phase5.west);
  \draw [arrow0] (phase5) -- (phase6);

  % Subtask arrows
  \draw [arrow] (p1t1) -- (p1t2);
  \draw [arrow] (p1t2) -- (p1t3);
  \draw [arrow] (p1t3) -- (p1t4);

  \draw[arrow] (p2t1.south east) -- (p2t4.north west);
  \draw[arrow] (p2t2) -- (p2t4);
  \draw[arrow] (p2t3) -- (p2t4);

  \draw [arrow] (p3t1) -- (p3t2);
  \draw [arrow] (p3t2) -- (p3t3);
  \draw [arrow] (p3t3) -- (p3t4);

  \draw [arrow] (p4t1) -- (p4t2);
  \draw [arrow] (p4t2) -- (p4t3);
  \draw [arrow] (p4t3) -- (p4t4);

  \draw [arrow] (p5t1) -- (p5t2);
  \draw [arrow] (p5t2) -- (p5t3);
  \draw [arrow] (p5t3) -- (p5t4);

\draw [arrow] (p6t1) -- (p6t2);
  \draw [arrow] (p6t2) -- (p6t3);
  \draw [arrow] (p6t3) -- (p6t4);

\end{tikzpicture}
\endgroup

\vspace{-0.5\baselineskip}

\caption{An overview of our proposed six-phase framework for developing, demonstrating, and evaluating a multimodal manufacturing-safety chatbot.}
\label{fig:method_overview}
\end{figure*}
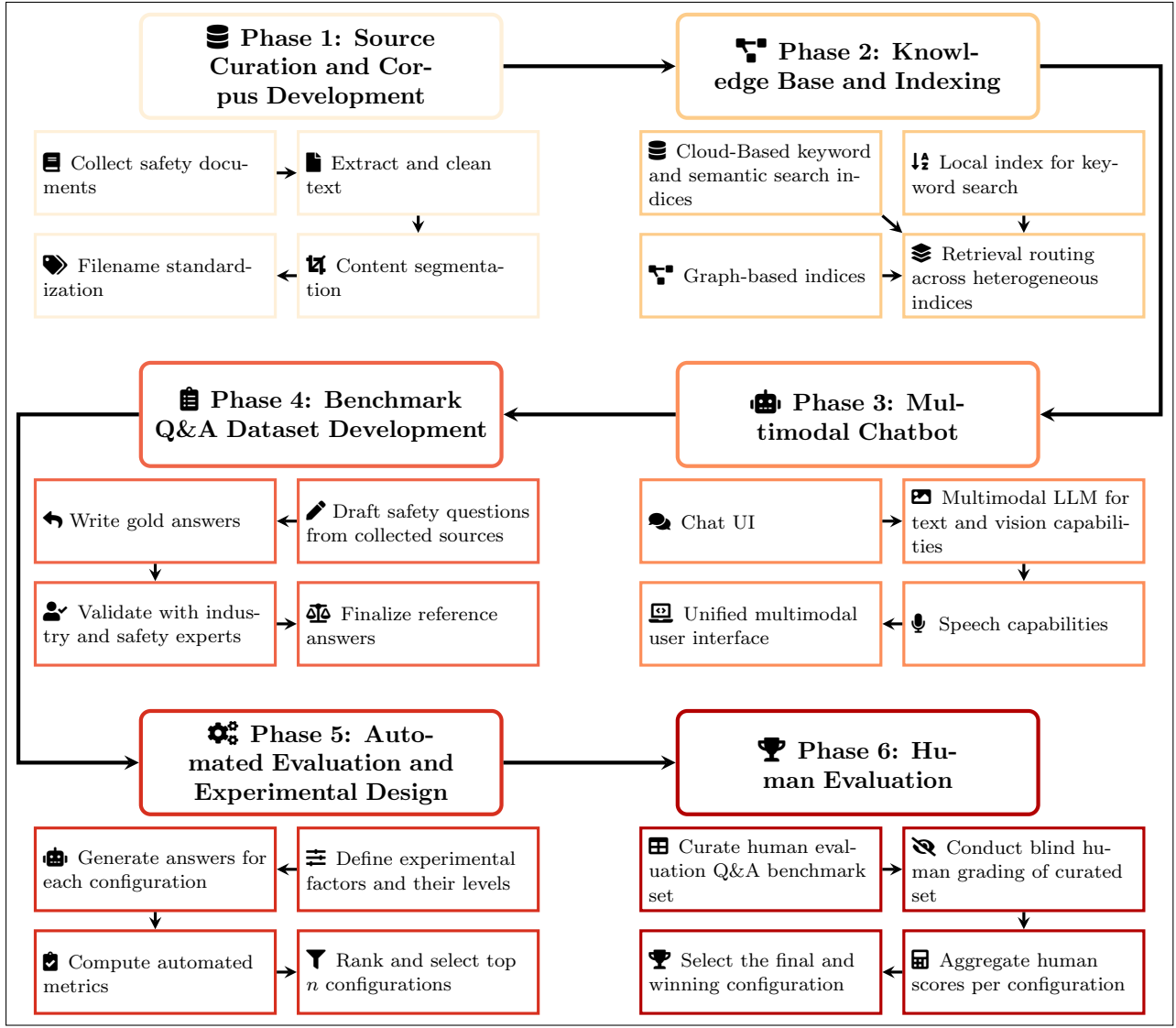

In the following sections, we design, demonstrate, and evaluate an artifact that satisfies the above design requirements. We do so by proposing the six-phase framework in Figure~\ref{fig:method_overview} for building manufacturing-safety chatbots grounded in expert-curated operational and safety documents. These phases align perfectly with our DSR methodology.
For example, Phases 1 and 2 are part of the third DSR stage called ``artifact design and development.'' In particular, in Phase 1, we curate and preprocess safety materials from OSHA, the National Institute for Occupational Safety and Health (NIOSH), and Original Equipment Manufacturer (OEM) manuals to build a reliable, structured corpus. In Phase 2, we construct multiple knowledge bases that integrate keyword, semantic, and graph indices to support flexible information retrieval for question answering (Q\&A). In Phase 3, we transition to the fourth design science stage, namely artifact demonstration. In particular, we demonstrate how a grounded, multimodal chatbot can be developed based on the knowledge base and indexing from the previous phase. Phases 4 to 6 are linked to the fifth stage of our DSR methodology, namely artifact evaluation. Phase 4 develops a benchmark Q\&A dataset, with expert-validated questions and reference answers. We evaluate multiple RAG configurations using automated metrics in Phase 5 to identify a reasonable subset of configurations for human evaluation. In Phase 6, we employ blind human evaluation to select the ``best'' configuration. The evaluations in Phases 4 to 6 enable us to implement a single configuration and deploy a multimodal chatbot that integrates text, visual, and speech interactions.
% , ensuring that its responses remain grounded in our curated source documents.

\section{Artifact Design and Development}
\label{sec:artifact_design}
In the third stage of our DSR methodology, we designed and developed a concrete artifact.  While many digital solutions could theoretically address the challenge of delivering accessible, accurate, and adaptive safety training in manufacturing, few can simultaneously meet the requirements of accuracy, low latency, and affordability identified in Section \ref{sec:methods}. For example, generic e-learning platforms or static digital manuals often fall short because they lack contextual awareness, interactivity, and real-time responsiveness. In contrast, modern AI technologies, such as a multimodal chatbot capable of processing text, images, and speech, offer a uniquely flexible and human-centered medium for safety communication. In particular, by embedding RAG techniques within a multimodal interface, such a system can ground its responses in authoritative safety documents while allowing workers to query information naturally through the mode most convenient to their environment. 
\hl{Given the advantages of conversational accessibility, multimodal capability, and ability to ground responses in authoritative documents, a RAG-based chatbot is a suitable artifact to operationalize the design requirements and realize the goals of Industry 5.0-aligned, human-centric safety training.}
We next describe the first two phases of our framework (Figure \ref{fig:chatbot}), which focus on data curation and the construction of RAG methods.

\subsection{Phase 1: Source Curation and Corpus Development} 
\label{subsec:Phase 1}

Phase 1 establishes the foundation of the chatbot development framework by focusing on the systematic curation and preparation of safety-critical source materials. This stage ensures that all subsequent modeling and retrieval processes are grounded in authoritative, high-quality data. 
% It involves collecting and organizing diverse safety documents into a unified corpus suitable for RAG. This phase includes four steps: (1) collecting the source documents, (2) extracting and cleaning text, (3) segmenting the content into meaningful units, and (4) standardizing how each segment is saved.

\subsubsection{Collect Safety Documents} 

\begin{figure}[htb!]
\centering

\usetikzlibrary{positioning, shapes, backgrounds}
\definecolor{color1}{RGB}{253,204,138}
\definecolor{color2}{RGB}{179,0,0}

\tikzset{
  phase/.style={
    rectangle, rounded corners, thick,
    draw=color1, fill=color1!12,
    text width=0.4\paperwidth,
    minimum height=0.9cm,
    align=center,
    font=\small
  },
  subtask/.style={
    rectangle, rounded corners,
    draw=color2, fill=white,
    text width=0.4\paperwidth,
    inner sep=4pt,
    align=left,
    font=\footnotesize
  }
}

% ---------- FIGURE CONTENT ----------
\begin{tikzpicture}[framed]

\node[phase] (title) {\textbf{Phase 1: Source Curation and Corpus Development}};

\node[subtask, below=0.25cm of title] (oshar) 
  {{\faBalanceScale}\ \textbf{OSHA Regulations (29 CFR 1910)}:\newline  
   \textcolor{white}{wsp}1910.147 (Control of Hazardous Energy: Lockout/Tagout),\newline 
   \textcolor{white}{wsp}1910.178 (Powered Industrial Trucks),\newline 
   \textcolor{white}{wsp}1910.211--1910.219 (Guarding to Mech. Power Transmission)};

\node[subtask, below=0.25cm of oshar] (oshat) 
  {{\faFile}\,  \textbf{OSHA Technical Documents}:\newline 
   \textcolor{white}{wsp}3120 (Control of Hazardous Energy Lockout-Tagout),\newline
   \textcolor{white}{wsp}3170 (Safeguarding Equipment \& Protecting from Amputations),\newline
   \textcolor{white}{wsp}OSHA Technical Manual (Sec. IV: Ch. 4 Industrial Robot Safety)};

\node[subtask, below=0.25cm of oshat] (niosh) 
  {{\faExclamationTriangle}\ \textbf{NIOSH Safety Alerts and Guidance}:\newline 
   \textcolor{white}{wsp}2011-156 (Lockout/Tagout to Prevent Injury during Maintenance),\newline 
   \textcolor{white}{wsp}85-103 (Preventing the Injury of Workers by Robots)};

\node[subtask, below=0.25cm of niosh] (oem) 
  {{\faTools}\ \textbf{OEM Manuals}:\newline  
   \textcolor{white}{wsp}Bridgeport Milling Machine (M-508, 2018),\newline
   \textcolor{white}{wsp}Haas Lathe Operator's Manual (96-8910, Rev. J),\newline 
   \textcolor{white}{wsp}UR5e Universal Robots User Manual};
\end{tikzpicture}

\vspace{-0.5\baselineskip}

\caption{Regulatory, safety, and OEM documents forming our safety corpus.} %The UR5e manual is highlighted in red, as we limit our evaluation procedure to queries within this document to remain within the conference’s page limits.}
\label{fig:phase1_sources}

\vspace{-\baselineskip}
\label{fig:source}

\end{figure}
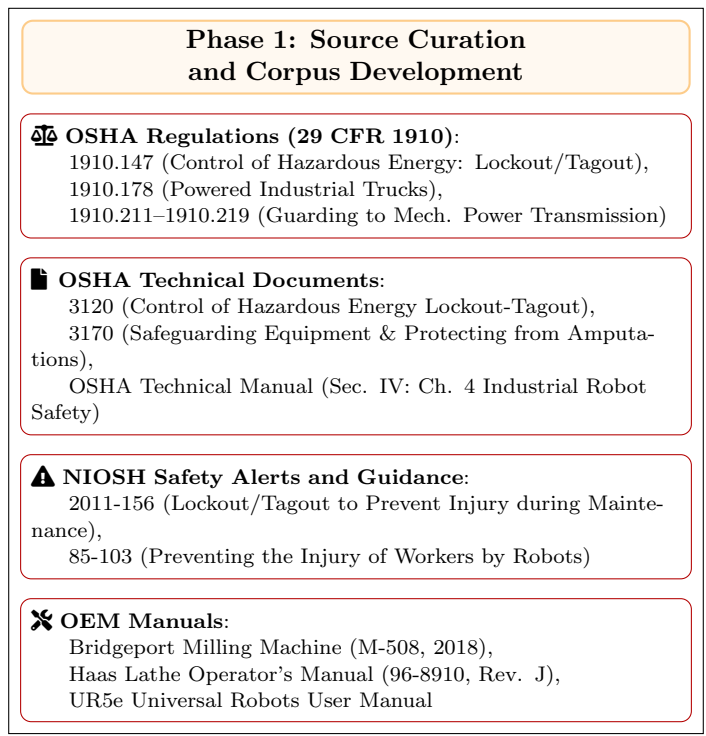

Grounding LLM responses in authoritative documents reduces hallucinations and improves factual correctness when compared to generation from the LLM alone \citep{lewis2020rag}.  We curated four categories of documents to ensure that the chatbot's responses are grounded in authoritative safety expectations: (1) OSHA Laws and Regulations, i.e., CFR 1910 Subparts O and N,  (2) OSHA technical guidance documents, (3) NIOSH safety alerts, and (4) OEM manuals for our target machines. Figure \ref{fig:source} summarizes our data sources.

\subsubsection{Extract and Clean Text} 

For each document, we first determined whether it was machine-readable. For machine-readable PDFs, we used the Python library \texttt{PyMuPDF}, which returns page-level text while preserving paragraph structure. If the document stored text as embedded images, we applied the Mistral OCR API to produce a text layer with page boundaries and paragraph breaks. After OCR, these documents were processed identically to machine-readable files to ensure a unified format across the corpus.

\subsubsection{Content Segmentation} 

For longer documents, such as OSHA standards and OEM manuals, we used the table of contents to identify the start of chapters and subchapters, and split the document accordingly. This segmentation keeps each text unit aligned with meaningful instructional topics. Shorter narrative documents, such as NIOSH safety alerts, often lack a usable table of contents. For these, we applied a length-based rule: documents exceeding 3,000 words were divided into smaller sections. This threshold balances retrieval performance by preventing segments from becoming excessively large or too fragmented.

\subsubsection{Filename Standardization} 

Once segmented, each section was saved as its own PDF, and any repeated header, footer, or boilerplate content was removed when present. We standardized filenames to include the source type, the document identifier, the section title, and the original page range. This naming convention ensures that each file is traceable back to its authoritative source and simplifies manual review, debugging, and alignment in downstream steps.

\subsection{Phase 2: Knowledge Base and Indexing} 

The goal of this phase is to construct the knowledge bases from the segmented documents in Phase 1 and develop the retrieval mechanisms that the chatbot relies on when answering questions. Prior research has shown that different retrieval strategies can surface different types of relevant content \citep{lewis2020rag}. Because manufacturing safety questions may use precise terminology or broader hazard descriptions, we examined multiple retrieval methods rather than relying on a single strategy. In later phases, we evaluate which approach performs best when answering the benchmark questions developed in Phase 4. 

\subsubsection{Cloud-Based Keyword and Semantic Search Indices}

We uploaded the segmented PDF documents from Phase 1 into an OpenAI-managed vector store, i.e., a database that stores documents as numerical representations (embeddings), enabling comparison and retrieval based on similarity in a high-dimensional vector space. OpenAI was selected as the vector store and retrieval provider for two  main reasons. First, its embedding models and retrieval API are tightly integrated, reducing engineering overhead and limiting potential sources of error in data preprocessing and indexing. Second, OpenAI provides a fully managed, scalable service with strong latency and reliability guarantees, simplifying deployment and supporting real-time interactive querying.

During ingestion, we specified a chunk size of 4,000 tokens (approximately 3,000 words; the default is 800 tokens) to ensure that the API did not further subdivide our segmented documents. \hl{As of Feb. 06, 2026}, the \texttt{OpenAI Retrieval API} supports two retrieval modes over the same stored data. In \emph{keyword search} mode, the system prioritizes exact or near-exact lexical matches between the query terms and the stored content. In \emph{semantic search} mode, the system rewrites or embeds the query and retrieves passages whose vector representations are close in meaning, even when they do not share surface-level wording. Keyword search is most effective when users know and reuse technical terms verbatim, whereas semantic search better supports users who describe their information needs in everyday language or with partial recall of the original phrasing.

\subsubsection{Local Index for Keyword Search} 

We constructed a local keyword-based index using the BM25 retriever in the Python \texttt{LangChain} library. BM25 employs probabilistic term weighting to rank documents based on estimated relevance and has remained a competitive baseline in information retrieval \citep{robertson2009probabilistic}. We selected BM25 for three primary reasons. First, as a purely lexical method, BM25 is well aligned with queries that contain distinctive regulatory language, component identifiers, or procedural step names, all of which tend to appear verbatim in technical manuals. Second, maintaining the BM25 index locally eliminates per-query API costs and reduces dependency on external services, which is advantageous for repeated experimentation and for potential deployment in cost-sensitive or bandwidth-constrained environments. Third, local execution provides consistently low latency and full control over the indexing and retrieval pipeline, facilitating reproducibility and fine-grained inspection of retrieval behavior.

\subsubsection{Graph-Based Indices} 

Graph-based retrieval methods allow the system to use both textual similarity and structural relationships among safety concepts, procedures, and component interactions. This is especially valuable in technical domains where relevant information spans multiple steps or subcomponents, and where safe outcomes depend on understanding dependencies rather than isolated passages. Following the \texttt{LangChain} \texttt{GraphRetriever} design \citep{langchain2025graph}, we implemented three graph RAG variants that differ in how they explore or constrain graph neighborhoods. \textit{AstraDB} serves as the online vector storage backend for all three configurations because it provides scalable embedding storage with native support for vector similarity search and integrates smoothly with graph-based retrieval workflows.

The three variants provide complementary strategies. The \texttt{Graph Eager} retriever expands outward through connected nodes to gather broader context, which helps when queries are imprecise. The \texttt{Graph MMR} variant applies maximal marginal relevance to balance relevance and diversity, reducing redundancy across similar segments. Finally, the vanilla similarity search baseline over \textit{AstraDB} performs traditional $k$-nearest-neighbor retrieval without graph traversal and serves as a reference for assessing the added value of graph methods.

\subsubsection{Retrieval Routing Across Heterogeneous Indices}
\label{subsubsection:router}

We implemented a unified retrieval router to coordinate access to the six retrieval mechanisms described above: (1) OpenAI keyword search, (2) OpenAI semantic search, (3) the local \texttt{BM25} retriever, (4) \texttt{Graph Eager}, (5) \texttt{Graph MMR}, and (6) vanilla similarity search over \textit{AstraDB}. The router decouples retrieval strategy selection from the rest of the conversational pipeline by standardizing how results are formatted and passed to the underlying LLM. This ensures that downstream evaluation compares retrieval strategies fairly and that performance differences can be attributed to retrieval behavior rather than formatting variations. Operationally, the routing layer allows retrieval strategies to be swapped in or out without modifying the chatbot architecture, enabling controlled experimentation in Phases 4 to 6 and simplifying demonstration in Phase 3.

\section{Artifact Demonstration}
\label{sec:demonstration}

The fourth stage of our DSR methodology is to ``\textit{demonstrate the use of the artifact to solve one or more instances of the problem}'' \citep{peffers2007design}. For our case, this involved deploying a publicly accessible multimodal chatbot interface that exposes the safety-assistance capabilities of the underlying RAG system to end users. The system focuses on three types of equipment: the Bridgeport Manual Mill, TL-1 CNC Machine, and the Universal  Robots Collaborative Robot (Cobot). The interface supports three input modes: (1) text-based dialogue, (2) image-based queries, and (3) speech-to-speech interaction.

Operationally, the fourth stage corresponds to Phase 3 of our development framework. We began by developing a text chat interface that allows workers to pose safety-related questions in plain language. 
The system then retrieves and synthesizes answers grounded in the authoritative documents curated earlier in the pipeline (see Section~\ref{subsec:Phase 1}). We next added an image input mode, enabling users to upload photographs of equipment, controls, or workspace conditions. 
The model interprets the visual context and issues a retrieval query tailored to the situation, returning relevant safety passages. 
Finally, we added a speech modality to enable hands-free questions and spoken responses.

The deployed system was implemented as a \texttt{Streamlit} application that integrates OpenAI’s \texttt{Responses} and \texttt{Realtime} APIs to provide text, image, and speech capabilities through a unified interface. Retrieval operates over the same corpus, vector stores, and indices established in earlier phases, preserving traceability between responses and their documentary sources. The application also exposes the retrieval router described in Section~\ref{subsubsection:router}, allowing users and researchers to select among different RAG configurations and to experiment with them in situ. This capability supports both practitioner use and systematic comparison in later evaluation phases.

\begin{figure}[htb!]
    \centering
    \includegraphics[width=0.4\paperwidth, frame]{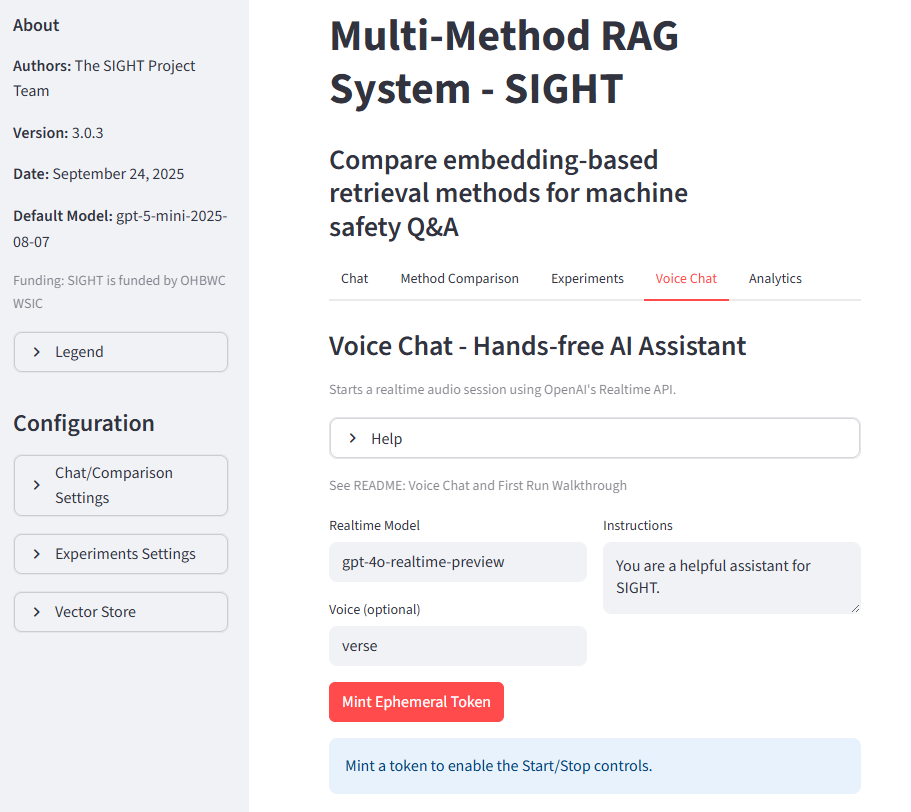}

    \vspace{-0.5\baselineskip}
    
    \caption{A screenshot of the voice module of our deployed chatbot. The chatbot is available at \url{https://sight.fsb.miamioh.edu/}.}
    \label{fig:chatbot}

    \vspace{-0.5\baselineskip}
\end{figure}

The system logs session-level analytics and provides transparent citations linking each answer to specific source passages, supporting explainability, auditability, and user trust. A screenshot of the deployed application is shown in Figure~\ref{fig:chatbot}, and the full code implementation is available in the project repository (see ``Data and Code Availability’’ section).

\section{Artifact Evaluation}
\label{sec:eval}
The fifth stage of our DSR methodology comprises ``\textit{comparing the objectives of a solution to actual observed results from use of the artifact in the demonstration}'' \citep{peffers2007design}. In our context, this stage corresponds to a systematic assessment of how well the proposed RAG configurations meet the design requirements identified earlier, namely high accuracy, low latency, and cost. To enable this assessment, we adopt a three-stage evaluation pipeline within our development framework. In Phase 4, we curate a domain-specific benchmark of safety-related queries and ground-truth reference answers derived from the authoritative corpus. In Phase 5, we use the above benchmark to conduct controlled experiments over multiple RAG configurations. Finally, in Phase 6, we conduct expert-based subjective evaluations to evaluate the most promising configurations, ultimately narrowing down to a single configuration that best balances performance, cost, and latency for deployment.

\subsection{Phase 4: Benchmark Q\&A Dataset Development}

We developed a comprehensive benchmark consisting of questions and gold-standard answers derived from the operational and safety manuals of the three target machines. This benchmark was designed to assess the chatbot’s ability to retrieve, reason about, and accurately convey safety-critical information across diverse levels of difficulty.

The process began with our research team conducting a detailed review of each machine’s official manuals. From these materials, we systematically extracted and formulated questions that represent realistic scenarios faced by operators and trainees in manufacturing settings. Each question was paired with a corresponding answer grounded explicitly in the manuals or regulatory guidelines, ensuring factual correctness and traceability. For instance, an example of an easy question was: \texttt{What is a pinch hazard in robot operation?} with the corresponding answer: \texttt{A pinch hazard occurs when body parts can be caught between moving robot parts or surfaces.} Easy questions can typically be answered from a single section of a manual. To better simulate realistic industrial information retrieval tasks, we also created more challenging questions that required integrating knowledge from multiple sections or documents. For example, a complex question might ask: \texttt{What are the external connection ports on the robot?} The answer should be \texttt{Teach Pendant Port, SD Card slot, Ethernet, USB 2.0 and USB 3.0, Mini Displayport, 10A Mini Blade Fuse}, which requires synthesizing details scattered across different technical sections. This ensured the benchmark tested direct retrieval and reasoning capabilities of the system.

After the initial Q\&A pairs were derived, we implemented a validation phase to ensure content accuracy and industrial relevance. Each Q\&A was reviewed either by experienced safety professionals from partner companies as well as by researchers with expertise in manufacturing, engineering technology, occupational safety and human factors engineering. Their feedback was used to refine the clarity, technical precision, and contextual relevance of the benchmark items. This iterative validation process guaranteed that the benchmark accurately represented the complexity of real-world safety knowledge and could be reliably used for evaluating AI system performance across different machines and hazard contexts. We refer to the final, validated answers in our benchmarks as \textit{gold-standard answers}.

In total, we created 60 questions for the \texttt{UR5 Cobot}, 51 for the Haas \texttt{TL-1 lathe}, and 42 for the \texttt{Bridgeport Manual Mill}. Due to space constraints, we henceforth focus on the \texttt{UR5 Cobot} in our evaluations since it exhibits greater operational complexity and a higher degree of human–machine interaction than the other two machines. We argue that the \texttt{UR5} benchmark sufficiently tests the system's ability to retrieve and synthesize information across multiple text segments (``chunks'') produced during corpus pre-processing.  As such, including additional results for other machines would provide little to no insight beyond what is demonstrated through the \texttt{Cobot} evaluation.
\hl{Our \texttt{UR5} dataset contains 23, 30, and 7 questions classified as easy, moderate, and challenging, respectively.}

\subsection{Phase 5: Automated Evaluation and Experimental Design} 

We used a full-factorial experimental design to study how retrieval method and model specification choices influenced the quality of answers generated by the AI. The purpose was to identify which configuration patterns consistently yielded correct responses obtained fast and cheaply before advancing a small number of candidate configurations for human evaluation in Phase 6. We vary three factors (see Table~\ref{tab:factors}), resulting in $24$ retrieval-generation \emph{pipelines}. Henceforth, we use the term pipeline to refer to a specific combination of the three factor levels used to generate an answer from the LLM/chatbot. Each pipeline is evaluated across all 60 benchmark questions, yielding $24 \times 60 = 1,440$ generated responses per evaluation metric.
% The full-factorial structure allows us to estimate main effects and selected two-factor interactions while treating higher-order interactions as residual variation, which aligns with our objective of identifying robust design patterns rather than optimizing configuration-specific fine-tuning.

\begin{table}[htb!]
\centering
\small

\vspace{-\baselineskip}

\caption{Factors and levels in the full-factorial experimental design.}
\label{tab:factors}
\begin{tabular}{p{0.09\textwidth} p{0.16\textwidth} p{0.18\textwidth}}
\toprule
\textbf{Factor} & \textbf{Purpose} & \textbf{Levels} \\
\midrule

\textbf{Retrieval}\newline \textbf{Approach}
& Determines how relevant safety text is selected & $\{$\texttt{OpenAI Keyword}, \newline  \texttt{OpenAI Semantic}, \texttt{BM25}, \texttt{Graph Eager}, \texttt{Graph MMR}, \; \texttt{Vanilla}$\}$ \\ \midrule

\textbf{LLM Used} & Controls model capacity and reasoning ability & $\{$\texttt{gpt-5-mini-2025-08-07},\newline \texttt{gpt-5-nano-2025-08-07}$\}$\\ \midrule

\textbf{Top}$-k$ \textbf{Retrieval Depth} & Number of chunks returned & $\{$3, \; 7$\}$ \\ \midrule

\multicolumn{3}{p{0.4\textwidth}}{\textbf{Total Pipelines} $= 6\times2\times1\times1\times2 = \, \mathbf{24}$}  \\
\bottomrule
\end{tabular}
\end{table}

For each pipeline, we generate an answer to every benchmark question using that pipeline's specific settings. 
All outputs are stored together with their pipeline, question, and replicate identifiers to support consistent evaluation. 
We used a single system prompt (set of instructions) in our experiments. 
\hl{The prompt explicitly instructs the model to use only the retrieved sources when generating answers, preventing it from relying on prior knowledge.
To promote concise and comparable outputs, the prompt instructs the model to provide direct answers in one or two sentences when possible and to add brief supporting sentences only when additional explanation is necessary.}
We set the underlying LLM's temperature parameter to 0 to ensure reproducibility, making responses deterministic for a given pipeline. 
\hl{Moreover, we set the maximum number of output tokens to $5,000$ to avoid truncation of safety-critical answers while still keeping costs bounded, and we set the reasoning effort  to $low$ to avoid introducing additional latency and cost.}

Given the generated answers from each pipeline alongside the gold-standard answers, we computed a set of evaluation metrics corresponding to the three design requirements in Section~\ref{sec:methods}. The complete set of metrics and their definitions are summarized in Table~\ref{tab:metrics}. First, accuracy was assessed using an LLM-as-judge that compares each generated answer to the gold-standard answer and returns a binary decision. The LLM-based judgment used standardized evaluation prompt templates provided in the LangChain ``RAG Evaluation'' tutorial \citep{langchain2025evaluate} and ChatGPT 5 as the model evaluator. Second, latency was measured as the end-to-end response time for answer generation. Third, cost values were built into the experiment design by restricting model selection to smaller/distilled versions of ChatGPT 5, namely \texttt{gpt-5-mini-2025-08-07} and \texttt{gpt-5-nano-2025-08-07}. These LLMs are considered to be high-performing and relatively low-cost. As of October 2025, the \texttt{gpt-5-mini-2025-08-07} was priced at \$0.25/1M input tokens and \$2/1M output tokens, and \texttt{gpt-5-nano-2025-08-07} was priced at \$0.050/1M input tokens and \$0.400/1M output tokens.

\begin{table}[htb!]
\centering
\small

\vspace{-0.5\baselineskip}

\caption{Summary of evaluation metrics used in our framework/experiment.}
\label{tab:metrics}

\begin{tabular}{p{.07\paperwidth} p{.29\paperwidth}}
\toprule
\textbf{Metric} & \textbf{Definition / Prompt Basis} \\
\midrule

Accuracy & Binary decision from LLM-as-judge comparing the generated answer with its corresponding gold-standard answer \\ \midrule

Latency & Model response time \\ \midrule

Cost & Token cost for \texttt{gpt-5-mini} and \texttt{gpt-5-nano} \\ %\midrule 

%Secondary & BLEU & $n$-gram precision with brevity penalty and smoothing (sentence-level BLEU) \\ \midrule

%Secondary & ROUGE-L & Longest common subsequence F-measure between generated and reference answers \\ \midrule

%Secondary & Cosine \newline Similarity & Cosine similarity $\left(\frac{X \cdot Y}{\|X\|\,\|Y\|}\right)$ computed between TF--IDF vectors of generated ($X$) and reference ($Y$) answers to measure semantic alignment independent of length \\ \midrule

%Secondary & Retrieval Relevance & LLM-as-judge indicating if retrieved passages relate to the question \\ \midrule

%Secondary & Groundedness (Faithfulness) & LLM-as-judge check that answer does not introduce information outside retrieved evidence \\ \midrule

%Secondary & Helpfulness & LLM-as-judge check that the answer directly and meaningfully addresses the question \\ \midrule
\bottomrule
\end{tabular}

\end{table}

\subsection{Phase 6: Human Evaluation} 
\label{sec:human_eval}

While automated evaluations provide valuable initial insights, they can only go so far in identifying the best retrieval–generation pipeline. For example, metrics such as LLM-as-judge may be imperfect for differentiating accuracy across solutions. To complement these results and select a single final pipeline, we conducted a human expert evaluation.

In the first stage, we evaluated all 24 pipelines on the full \texttt{UR5 Cobot} benchmark. Performance was assessed primarily on accuracy. Latency was considered as a secondary criterion, and cost was treated as an acceptability threshold rather than a ranking factor. When pipelines had no practical differences in accuracy or latency, we relied on their accuracy ranking to guide the selection of candidates. Based on this evaluation plan, we narrowed the original set of 24 pipelines down to 2.

\begin{figure}[htb!]
    \centering
    \includegraphics[width=0.4\paperwidth, frame]{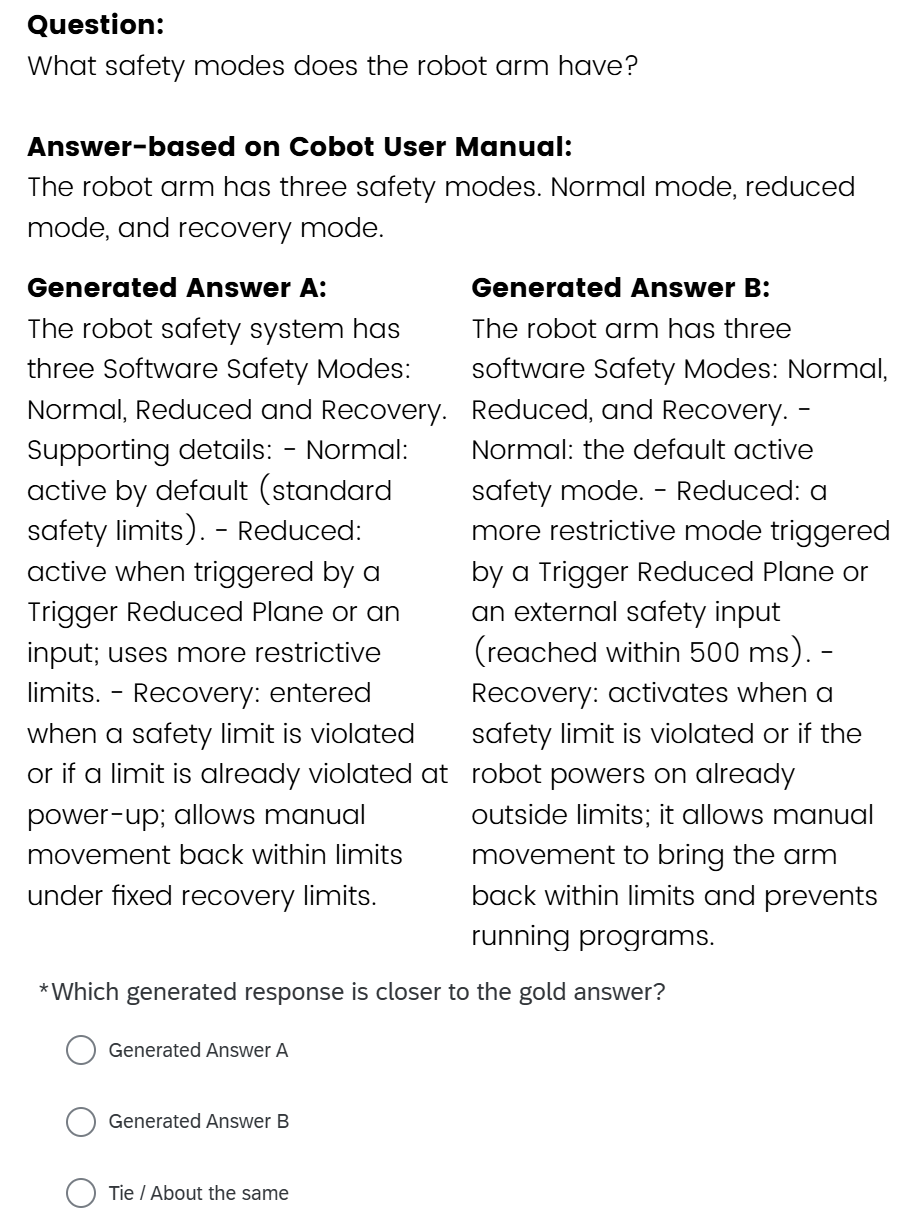} % I also uploaded a no color version under figs
    \caption{\hl{A screenshot of the interface used for human evaluation for blind side-by-side comparison of model outputs}.}
    \label{fig:qualtrics}
\end{figure}

After this narrowing step, we drew a random sample of ten questions from the \texttt{UR5 Cobot} benchmark for human evaluation. For each sampled question, we retrieved the corresponding answers generated by the two selected pipelines.
These ten questions and answers were then presented to expert evaluators through a purpose-built web interface that enabled blind side-by-side comparison of model outputs. Figure \ref{fig:qualtrics} shows this evaluation interface. This setup allowed us to distribute a structured survey to domain specialists in manufacturing, occupational safety, and human factors.
\hl{All evaluators were safety engineers and academic researchers; we did not include front-line operators in this study.
A planned future study includes assessing how performance and reliance change for less technically literate user groups, including new operators}.

\subsection{Results}

\begin{figure*}[htb]
    \centering
    \includegraphics[width=0.99\linewidth, frame]{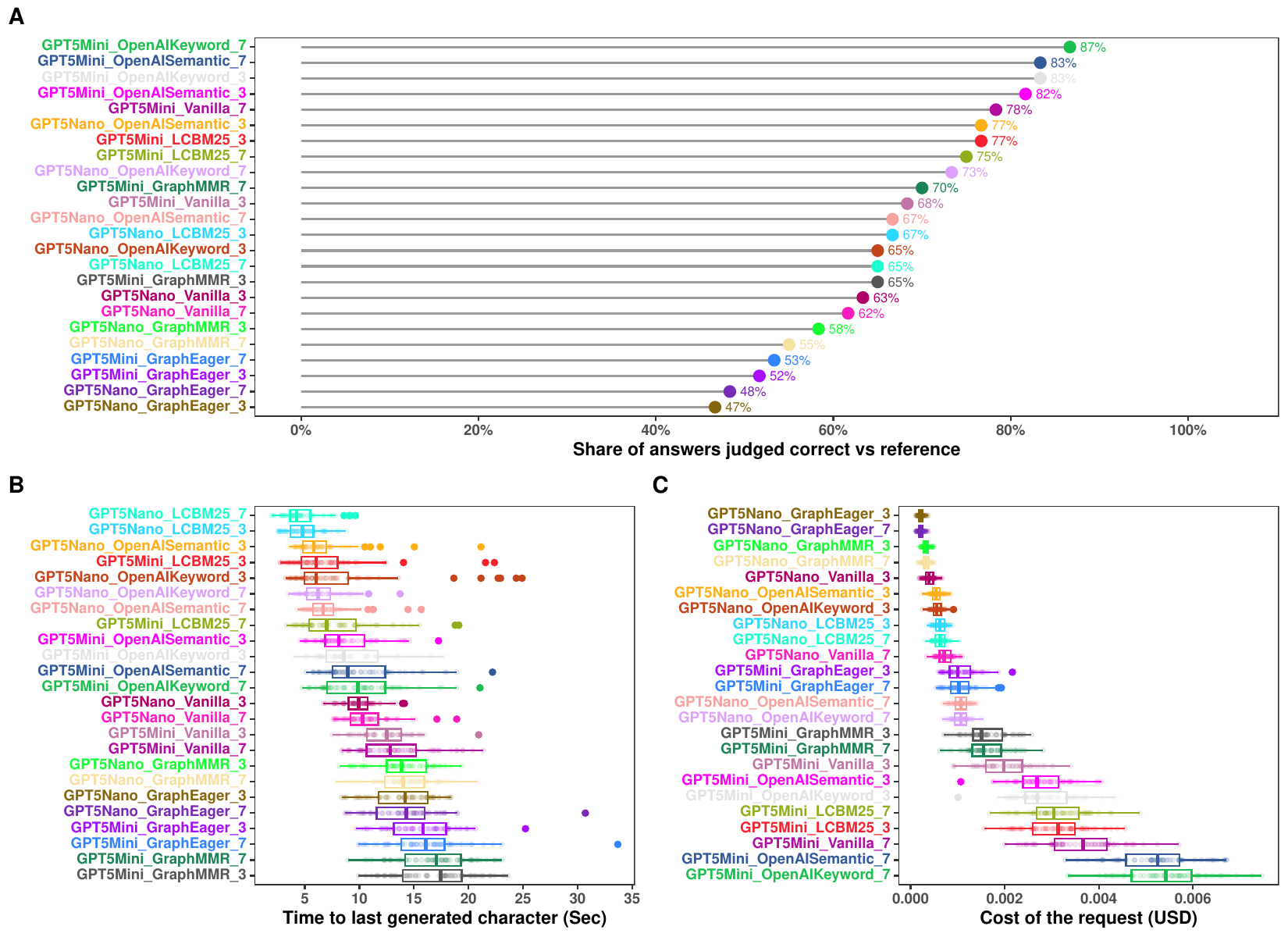} % I also uploaded a no color version under figs
    \caption{Comparison of the performance of our 24 RAG pipelines across correctness (A), generation time, i.e., latency (B), and cost (C). The y-axis in each subplot utilizes the \texttt{LLMModel\_RagApproach\_TopKChunksUsedAsInput} naming convention. For each subfigure, we sort the pipelines by performance, placing the best at the top (using median performance for the box plots). }
    \label{fig:eval_colored}
\end{figure*}

We next report our experimental results by comparing different RAG configurations to understand their impact on the metrics of interest. 
Specifically, we highlight key trends and trade-offs observed in our experiments, leading to the configuration that offers the most practical balance for deployment in manufacturing training settings.
Figure \ref{fig:eval_colored} summarizes our results.

\subsubsection{Accuracy}
\label{subsubsec:accuracy}

First, we examined how each of the pipelines affected the proportion of answers judged correct relative to the gold-standard answers (as depicted in Figure~\ref{fig:eval_colored}A). Our reported analyses in this subsection focus on main effects. That is, the average impact of each factor when collapsing over the others, rather than on specific interactions between model, retrieval depth, and retrieval method.

Across all conditions, the larger \texttt{gpt-5-mini-2025-08-07} model achieved a higher overall correctness rate ($0.73$) than the smaller \texttt{gpt-5-nano-2025-08-07} model ($0.62$). A paired McNemar test confirmed that this difference was statistically significant, $\chi^2(1, N = 720) = 19$, \ $p < .001$. In other words, the presumed additional capacity of the \texttt{mini} variant translated into more accurate safety-related answers in our benchmark when averaging across retrieval depths and approaches.

Next, we assessed the effect of the number of retrieved chunks provided to the LLM (\texttt{top k} = $3$ vs. $7$). Accuracy was similar across these conditions, with correctness proportions of $0.67$ and $0.68$ for \texttt{top k} = $3$ and \texttt{top k} = $7$, respectively. McNemar’s test found no statistically significant difference between these retrieval-depth settings, $\chi^2(1, \ N = 720) = 38$, \ $p = 0.445$. %This suggests that, within the range studied, increasing the number of retrieved chunks yielded only minimal gains in correctness for our tasks. 
From a practical perspective, correct proportions of $0.67$ and $0.68$ are also practically indistinguishable, which implies that the added computational cost of retrieving more chunks may not be justified in our case.

Finally, we investigated the main effect of retrieval approach on answer accuracy. Cochran's $Q$ test indicated a significant effect of retrieval method, $Q(5) = 103.24, \ p < 0.001$. Averaged across models and retrieval depths, the highest proportions of correct answers were observed for the \texttt{OpenAI Keyword} and \texttt{OpenAI Semantic} approaches (both $0.77$), followed by \texttt{BM25} ($0.71$), \texttt{Vanilla} ($0.68$), \texttt{Graph MMR} ($0.62$), and \texttt{Graph Eager} ($0.50$). Post-hoc pairwise McNemar tests with Holm adjustment, focusing on the three best-performing methods, revealed no significant differences among \texttt{OpenAI Keyword}, \texttt{OpenAI Semantic}, and \texttt{BM25} (all adjusted $p = 0.2$), indicating statistically equivalent performance at the top end. Both OpenAI-based methods, however, significantly outperformed all remaining baselines, i.e., \texttt{Graph Eager}, \texttt{Graph MMR}, and \texttt{Vanilla} (all adjusted $p < 0.05$), highlighting the advantage of OpenAI's retrieval strategies for maximizing response correctness.

These results characterize average (main) effects of each factor. That is, we ask questions such as ``Is \texttt{mini} more accurate than \texttt{nano} on average?'' or ``Which retrieval approach performs best on average?'' rather than examining the full pattern of interactions (for example, whether certain retrieval methods benefit \texttt{mini} more than \texttt{nano}, or whether \texttt{top k} = $7$ helps some methods but not others). Analyses at the level of specific model–method–depth combinations, e.g., those in Figure~\ref{fig:eval_colored}A, would be required to fully characterize such interaction effects.

\subsubsection{Latency}
\label{subsubsec:latency}

We next examined how each of the model choice, retrieval depth, and retrieval approaches influenced end-to-end response time (Figure \ref{fig:eval_colored}B). As expected, the smaller \texttt{gpt-5-nano-2025-08-07} model generated responses more quickly than the larger \texttt{gpt-5-mini-2025-08-07} model. Mean latency for the \texttt{mini} model was $M_{\text{mini}} = 12.03$\,s, compared with $M_{\text{nano}} = 9.53$\,s for the \texttt{nano} model. A one-sided paired $t$-test (testing whether the mini model was slower than the nano
model) confirmed that this difference was statistically significant ($t = 20.80$, \ $p < 0.001$). This means that the higher accuracy gains associated with the \texttt{mini} model (described in Section \ref{subsubsec:accuracy}) came with a marked increase in response time.

Retrieval depth also produced a measurable latency effect. Providing the LLM with $k = 7$ retrieved chunks yielded a mean response time of $M_{k=7} = 10.98$ seconds. On the other hand, providing $k=3$ retrieved chunks results in a mean response time of $M_{k=3} = 10.58$ seconds. Although the absolute difference was modest (approximately $0.4$\ seconds on average), a one-sided paired $t$-test (testing whether $k=7$ was slower than $k=3$) indicated that it was statistically significant ($t = 3.52$, \ $p < .001$). This means that increasing the number of retrieved chunks thus led to a small but measurable increase in latency. 

Finally, latency varied substantially across retrieval approaches. A Friedman test revealed a significant overall effect of retrieval method on response time ($\chi^2(5) = 822.53$, \ $p < 0.001$). Mean latencies (in seconds) for each approach were: $M_{BM25} = 6.11$, $M_{OpenAI Semantic} = 8.08$, $M_{OpenAI Keyword} = 8.56$, $M_{Vanilla} = 11.61$, $M_{Graph Eager} = 14.97$, and $M_{Graph MMR} = 15.35$. Post-hoc Wilcoxon signed-rank tests with Holm correction showed that all pairwise differences were statistically significant ($p_{\text{corrected}} < 0.05$) except for the comparison between \texttt{OpenAI Keyword} and \texttt{OpenAI Semantic} (adjusted $p = 0.71$), whose latencies were statistically equivalent.

\hl{It is important to note that our latency measurements reflect end-to-end time to generate the \emph{entire} response, rather than time-to-first-token.
As a result, these values can overestimate perceived latency in practice, since users typically begin reading or listening while the response is still being generated.
More importantly, the chatbot is intended for informational safety guidance (e.g., training and decision support) and is not designed for time-critical protective actions such as emergency stop activation or lock-out/tag-out execution, for which latencies on the order of seconds (or less) are required.}

\subsubsection{Cost}

To complement the accuracy and latency analyses, we evaluated how each of the model size, retrieval depth, and retrieval method affected the monetary cost per request (Figure \ref{fig:eval_colored}C). Costs were computed from input and output token usage under the October 2025 pricing for \texttt{gpt-5-mini-2025-08-07} and \texttt{gpt-5-nano-2025-08-07}.

Averaged across retrieval settings, the \texttt{mini} model was more expensive than the \texttt{nano} model. The \texttt{mini} variant cost $\$0.0028$ per query on average, compared with $\$0.00056$ for the texttt{nano} model. A one-sided paired $t$-test confirmed that this difference was statistically significant ($t = 48.67$, $p < 0.001$), confirming that the additional power of the \texttt{mini} model comes with a substantially higher token cost.

Second, we assessed the impact of retrieval depth. Pipelines that used $k=3$ retrieved chunks had a mean per-query cost of $\$0.0013$, compared with $\$0.002$ for $k=7$. A one-sided paired $t$-test (testing whether $k=7$ was more expensive than $k=3$) indicated that this difference was statistically significant ($t = 18.25$, \ $p < 0.001$). Thus, on average, deeper retrieval nearly doubled the per-query cost, slightly increased the latency (Section \ref{subsubsec:latency}) and produced modest gains in accuracy (Section \ref{subsubsec:accuracy}).

Finally, we investigated cost differences across retrieval approaches, averaging over models and retrieval depths. A Friedman test showed a significant overall effect of retrieval method on cost ($\chi^{2}(5) = 942.05$, \ $p < 0.001$). Mean per-query costs were lowest for \texttt{Graph Eager} ($\$0.00064$), followed by \texttt{Graph MMR} ($\$0.00097$), \texttt{Vanilla} ($\$0.00168$), \texttt{BM25} ($\$0.00185$), \texttt{OpenAI Semantic} ($\$0.00240$), and \texttt{OpenAI Keyword} ($\$0.00247$). Post-hoc pairwise Wilcoxon signed-rank tests with Holm correction show that all methods differed significantly from one another (all adjusted $p < 0.05$). 

Although the cost differences across methods were statistically significant, their practical impact is modest in our anticipated deployment setting. Even the most expensive configurations differ by only a few thousandths of a dollar per query. At an expected volume of fewer than 10,000 queries per month, these differences amount to only a few dollars in monthly spend. Thus, while statistically significant, the observed cost effects are not likely to be decisive for chatbot usage at our scale.

\subsection{Best Pipeline Selection}

Selecting a single RAG configuration for deployment requires balancing the competing objectives of accuracy, latency, and cost. Our experimental findings indicate that no single pipeline dominates across all three criteria, reinforcing the inherently multi-objective nature of RAG system optimization. Out of the 24 pipelines, the OpenAI keyword retrieval method paired with \texttt{gpt-5-mini-2025-08-07} consistently produced the highest accuracy scores, reaching 86.66\% and 83.33\%, respectively, when using retrieval depths of $k = 7$. The average latency and cost values for both pipelines were approximately 10 seconds and $\$0.005$, respectively.

Subsequently, we followed the evaluation protocol described in Section \ref{sec:human_eval} and conducted a blind human-expert assessment of the top-performing pipelines. Eleven manufacturing and occupational safety experts, including seven external to our research team, participated in the review. They evaluated responses from the two most accurate pipelines (\texttt{gpt-5-mini paired} with \texttt{OpenAI Keyword} and \texttt{OpenAI Semantic} retrieval at $k$ = 7) using a randomly selected subset of ten benchmark questions. Across the resulting 110 blinded comparisons ($11 \text{ experts } \times \, 10 \text{ questions}$), the experts preferred the the \texttt{OpenAI Keyword} retrieval in 45 (41\%) cases, the \texttt{OpenAI Semantic} retrieval in 29 (26\%) cases, and judged 36 (33\%) cases as ties. Taken together, the expert evaluations corroborated the LLM-as-Judge evaluation approach and decisively supported the \texttt{gpt-5-mini-2025-08-07} + \texttt{OpenAI Keyword} + \texttt{top-k = 7} pipeline as the most suitable configuration for deployment. Although this configuration is not the cheapest nor the fastest, it offers the best trade-off for safety-critical environments. It  maximizes retrieval and generative accuracy (the system's most important requirement), while maintaining acceptable levels of cost and latency.

\section{Concluding Remarks}
\label{sec:conc}
This study introduced a multimodal, RAG-based chatbot framework designed to advance safety training and hazard recognition within Industry 5.0 manufacturing environments. 
Grounded in a rigorous Design Science Research methodology, the work systematically addressed three critical requirements for next-generation safety systems, namely, accuracy, latency, and cost, through an experimentally validated approach.
The final  system integrates text, image, and voice modalities with multiple retrieval strategies to deliver contextually grounded, real-time safety guidance. 
Beyond the artifact, a secondary contribution of our work is a publicly available benchmark that can help to evaluate future AI manufacturing solutions.

We argue that our empirical results are of great value to practitioners, as they demonstrate how different retrieval strategies and model configurations can exert a significant influence on system performance.
In particular, our findings highlight the multi-objective nature of RAG configuration optimization, where trade-offs among correctness, speed, and expense must be balanced. 
This leads to interesting future research directions, where one can explore Pareto-based optimization frameworks to systematically identify non-dominated solutions that best balance accuracy, latency, and cost given a manufacturing facility’s needs and budget. 
Our results with a local RAG approach suggest another promising avenue for future work. 
In particular, by fine-tuning lightweight local LLMs and combining them with efficient, on-premise RAG pipelines, one can potentially reduce both operational costs and response latency while preserving accuracy. 
\hl{Finally, given the safety-critical context, future work should evaluate trust calibration by measuring over-reliance (accepting incorrect advice) and under-reliance (ignoring correct guidance) in realistic user studies. Such studies can also test behavioral interventions to better align reliance with system reliability.}
These directions can further enhance our contribution of a framework and empirical foundation for next-generation AI-driven safety training systems, which marks a step toward human–AI collaboration that enhances safety, efficiency, and resilience in smart manufacturing environments.

%LIMITATIONS:
%latency: hard to measure on cloud-based systems since they are dynamic. We are measuring all tokens, as opposed to first one.

%summarize DR. For accuracy, say Accuracy is not merely a technical goal but an ethical imperative: workers must be able to trust that what they learn reflects the true and safe way to act within their operational context. For latency, say, responsiveness transforms the system from a passive educational tool into an active, real-time safety partner. For cost, say By reducing financial barriers to adoption, this requirement ensures that advanced safety training remains equitable, sustainable, and scalable across diverse industrial contexts.

% Note: when evaluating artifact agasint accuracy mention that ``instructional content must be validated against authoritative safety documentation and reviewed by domain experts prior to delivery to ensure factual precision and user trust, while avoiding improper actions and serious injury" , and consistent with official standards and manufacturer guidelines.

\section*{Data and Code Availability}
Code and evaluation benchmarks are available at \url{https://github.com/fmegahed/safety_rag_evaluation}.

\section*{Acknowledgments}
This work was funded by the Ohio Bureau of Workers' Compensation Worker Safety Innovation Center (WSIC) Grant \#\texttt{WSIC26-250206-009}. Dr. Megahed and Carvalho also received support from the \textit{Raymond E. Glos Professorship} and the \textit{Dinesh \& Ila Paliwal Innovation Chair}. The authors acknowledge the \textit{Farmer School of Business} for hosting our chatbot. The authors also appreciate the valuable feedback provided by \textit{Engineered Profiles}, \textit{Yamaha Motor Company of North America}, and \textit{MeetKai, Inc.} on our benchmark Q\&A datasets.

\bibliographystyle{elsarticle-num} 
\bibliography{refs}

\end{document}